\documentclass[nofootinbib]{revtex4}

\usepackage{amsmath,amssymb,amsfonts}
\usepackage{hhline}
\usepackage[colorlinks]{hyperref} 
\usepackage[pdftex]{graphicx}
\graphicspath{{./figs/}}

\begin{document}

\newcommand{\bra}[1]{\big\langle#1\big\vert}
\newcommand{\ket}[1]{\big\vert#1\big\rangle}
\newcommand{\tr}{\mathrm{Tr}}

\title{Quantum Walks of SU(2)$_k$ Anyons on a Ladder}
\author{Lauri Lehman}
\author{Gavin K. Brennen}
\affiliation{Centre for Engineered Quantum Systems (EQuS), Macquarie University, Sydney, NSW 2109, Australia}
\author{Demosthenes Ellinas}
\affiliation{Technical University of Crete, Department of Sciences, M$\Phi$Q Research Unit, GR-731 00 Chania, Crete Greece}

\begin{abstract}
We study the effects of braiding interactions on single
anyon dynamics using a quantum walk model on a quasi-1-dimensional ladder filled with stationary anyons.  
The model includes loss of information of the coin and nonlocal fusion degrees of freedom on every second time step, such that the entanglement between the position states and the exponentially growing auxiliary degrees of freedom is lost.  The computational complexity of numerical calculations reduces drastically from the fully coherent
anyonic quantum walk model, allowing for relatively long simulations for anyons which are spin$-1/2$ irreps of SU(2)$_k$ Chern-Simons theory.
We find that for Abelian anyons, the walk retains the ballistic spreading velocity just like particles
with trivial braiding statistics. For non-Abelian anyons, the numerical results indicate that the
spreading velocity is linearly dependent on the number of time steps.  By approximating the Kraus generators of the time
evolution map by circulant matrices, it is shown that the spatial probability distribution for the $k=2$ walk, corresponding to Ising model anyons, is equal to the classical unbiased random walk distribution.
\end{abstract}

\maketitle

\section{Introduction}
Quantum walks have proven to be a fruitful model to study the dynamics of particles on lattices or more general graphs.  The effects of exchange statistics on multiparticle quantum walks has been studied for bosons and fermions with a notable effect of statistics.  For example it has been shown that quantum walks with non-interacting bosons give rise to effective interactions which lead to Bose-Einstein condensation in graphs with spatial dimension $d<2$ \cite{Vezzani}.  Quantum walks with pairs of bosons or fermions have been studied in the context of the graph isomorphism problem by relating the Green's function of the evolution to the spectrum of the graph.  It was shown that these models have much more computational power when the particles are allowed to interact because they can distinguish non-isomorphic graphs that noninteracting particles cannot \cite{Coppersmith}. Recently using integrated photonics there has been experimental realization of two particle quantum walks which simulated bosonic, fermionic, and semionic (a type of Abelian anyon) statistics \cite{Sansoni}.   

A natural extension of multiparticle quantum walks is to consider anyons which are particles that exist in two dimensions and have richer exchange statistics than bosons or fermions.  In the case of Abelian anyons, exchanges, or braids, yield some model dependent phase $e^{i\phi}$ while for non-Abelian anyons there is a large Hilbert space of fusion degrees of freedom of the multi particle system and a braid performs a unitary transformation on this vector space.  A quasi-1-dimensional model for anyonic quantum walks was introduced in Ref. \cite{Wang2010} where it was shown that Abelian anyons have quadratic variance just like standard quantum walks with trivial statistics but that non-Abelian anyons create entanglement during the walk purely due to braiding that acts to decohere the spatial degree of freedom. In Ref. \cite{isingaqw} the quantum walk on a quasi-1-dimensional ladder was studied in detail for one type of non-Abelian anyon known as the Ising model anyon.  This anyon corresponds to a spin$-1/2$ irrep. of   $SU(2)_{k=2}$ Chern-Simons theory, and the variance of the probability distribution of the walker was
proven to be asymptotically linear in time  with coefficient one,  corresponding to the classical random walk behaviour.  That particular case points out that there is a fundamental difference between the dynamics of Abelian vs. non-Abelian anyons.

In this work we explore these differences for more general anyon models,
specifically for spin$-1/2$ irreps of all values of the index $k$ in $SU(2)_k$ Chern-Simons theory. 
Numerical results show that for small number of time steps, the position distribution approaches the quantum walk distribution as index $k$ grows \cite{Wang2010}, and it was conjectured that the distribution looks classical
when $k\ll t$. The problem with the fully coherent anyonic quantum walk is that to evaluate the probability
distribution for $t$ iterations of the discrete evolution operator, the number of paths which contribute
to  the walk grows exponentially with $t$.  It turns out that evaluating each path is equivalent to computing the Jones polynomial for the link corresponding to the closed world line of that path, a computation which is itself exponentially hard (except for the cases $k=1,2,4$ \cite{Jaeger}) in the number of time steps $t$. To simplify the calculations for general $k$, we adopt here a special quantum walk protocol
which introduces loss of memory to the coin and fusion degrees of freedom. Such a protocol
allows for calculating the walk distribution for a relatively large number of time steps and to use certain approximations
to diagonalize the walk operator and obtain analytical results.


\section{Anyonic $V^2$ quantum walk}

\begin{figure}[h]
\includegraphics[width=.6\textwidth]{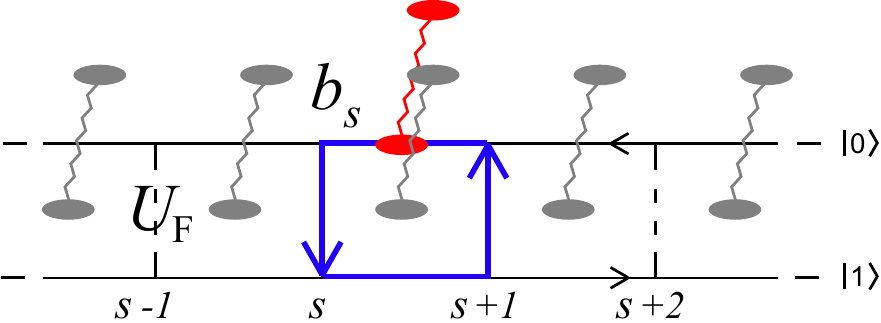}
\caption{The anyonic quantum walk on a ladder of point contacts. The red oval is the walker anyon which
moves along the black lines, and the grey ovals are the stationary anyons which occupy the islands between the point contacts.  The anyons are initialized in the state $\ket{\Phi_0}$ meaning they are created out of the vacuum in pairs as indicated by the ovals joined by strings, and half of each pair is outside the ladder not participating in the walk.
When on the upper(lower) edge the walker always to the left(right).
The occupation sites of the mobile anyon are labelled by the positions $s$ of the
point contacts where the mobile anyon can tunnel. The tunneling matrix between the legs at each rung corresponds to the coin shuffling operator $U_{\text{F}}$.
Each lattice shift corresponds to a braid between the worldlines of the
walker anyon and a stationary anyon, represented by the generator $b_s$.}
\label{fig:anyonwalk}
\end{figure}

The anyonic quantum walk is a dynamical model for a single anyon (mobile or ``walker" anyon) moving
in a lattice of fixed anyons (stationary ``background" anyons). The mobile anyon has an extra
degree of freedom called the \emph{coin} which is coupled to the direction of its movement.
Here we consider a quasi-1-dimensional lattice of background anyons where the walker has only
two directions of movement, left or right, and the coin is thus two-dimensional.  Although the anyons
are arranged on a line, they exist on a two-dimensional manifold such that the mobile anyon can
pass them from above or below without coming into contact with the stationary anyons.  This dynamics is accomodated by a walk on a ladder where the top(bottom) leg corresponds to mode $\ket{0}$$(\ket{1})$ of the coin, the sites correspond to the rungs and are labelled by an integer $s$, the stationary anyons are pinned in the islands of the ladder (see Fig. \ref{fig:anyonwalk}).

The arrangement in Fig. \ref{fig:anyonwalk} is a crude model of quasiparticle dynamics in
Fractional Quantum Hall samples which support anyonic quasiparticle excitations
\cite{Willett,Bonderson}. The edge modes
(horizontal in Fig. \ref{fig:anyonwalk}) carry mobile quasiparticles with a current direction determined by an externally applied voltage, and bulk quasiparticles
can be pinned at antidots within the islands.  
The upper and lower edge modes are mixed by introducing point contacts controlled by front gates  between the islands,
allowing tunneling from edge to edge. The coin is thus encoded as propagation to left or right
on the upper or lower edges respectively, and the scattering matrix that describes the tunneling
between the edges represents the quantum coin flip operator. The model assumes a hardcore
approximation for the anyons, such that they are always far apart, no polarization to
fusion channels occurs and the anyons interact only non-locally via their braiding statistics.
The model is also chiral: the particles on the upper edge are only allowed to move left
and on the lower edge only to the right, and all braids are therefore anti-clockwise.

When the walker shifts
between lattice sites, the wave function is multiplied by the braid generator
that corresponds to braiding the walker and the stationary anyon. The total Hilbert space
is the tensor product space of the spatial sites, fusion degrees of freedom and the coin,
$\mathcal{H}=\mathcal{H}_\text{space}\otimes\mathcal{H}_\text{fusion}\otimes\mathcal{H}_\text{coin}$.
The position space is spanned by $\big\{\ket{s}\big\}_{s=0}^{N-1}$, the fusion space accommodates $2N$
anyons ($N-1$ stationary anyons, the walker anyon and their antiparticle pairs) with total charge
vacuum, and the coin space is spanned by $\big\{\ket{0},\ket{1}\big\}$. We assume periodic boundary
conditions, and choose the number of sites to be always larger than the length of the walk,
$N\geq4t+1$. The walker starts initially from the localized state $\ket{s_0}\bra{s_0}$. For reasons described below, rather than fully coherently evolving the system over many steps, the walk evolves coherently for two cycles followed by tracing over fusion and coin degrees of freedom. Afterward the fusion state is reset to the vacuum pair state $\ket{\Phi_0}\bra{\Phi_0}$, the coin state to  $\ket{c_0}\bra{c_0}$, and the dynamics is iterated:
\begin{equation} \label{eqn:aqwevolution1}
\rho_{\text{S}}(t+1)=\tr_{\text{f,c}}\big(\big(U_{\text{S}}U_{\text{F}}\big)^2\:
\rho_{\text{S}}(t)\otimes\ket{\Phi_0}\bra{\Phi_0}\otimes\ket{c_0}\bra{c_0}\:
\big((U_{\text{S}}U_{\text{F}})^2\big)^\dag\big)
\end{equation}
\begin{equation}
U_\text{F}=I_\text{space}\otimes I_\text{fusion}\otimes H
\end{equation}
\begin{equation}
H=\frac{1}{\sqrt{2}}\begin{pmatrix}1&1\\1&-1\end{pmatrix}
\end{equation}
\begin{equation} \label{eqn:aqwevolution4}
U_\text{S}=\sum\limits_{s=0}^{N-1}\Big(S_{s+1}^-\otimes b_{s}\otimes P_0
+S_s^+\otimes b_{s}\otimes P_1\Big)
\end{equation}
when the Hadamard matrix $H$ is chosen as the coin flip operator,
$S_s^-=\ket{s-1}\bra{s}$ and $S_s^+=\ket{s+1}\bra{s}$ are the shift operators and
$P_0=\ket{0}\bra{0},\;P_1=\ket{1}\bra{1}$ are projectors to the coin states.
The braid generators $b_s$ are the matrices associated with braiding the anyons
$s$ and $s+1$ in anticlockwise direction. Due to the periodic boundary conditions,
there is also a braid generator $b_{N-1}$ associated
with braiding the anyons 1 and $N$ which can be expressed in terms of the other generators%
\footnote{This generator can be explicitly written as
$b_{N-1}b_{N-2}\ldots b_2b_1b_2^\dag b_3^\dag\ldots b_{N-2}^\dag b_{N-1}^\dag$},
but we choose the system size so that the walker never crosses the boundary and this
generator is never applied.   A string of braid generators,
$B=\prod_{s_j} b_{s_j}$ is called a \emph{braid word}. The braid matrices are determined by the anyon model
in question. In a physical setting, the properties of the underlying material supporting
the anyonic excitations determine the anyon model. 

We consider a general class of non-Abelian
anyons corresponding to spin$-1/2$ irreps of the quantum group SU(2)$_k$ which is a group with a deformed version of the
SU(2) algebra. The fusion rules of the SU(2)$_k$ model satisfy the triangle inequality and integer sum condition as in SU(2):
\[ j_1\times j_2=\sum\limits_{j=|j_1-j_2|}^{j_1+j_2}j \]
 but
with two restrictions on the total spin charge $j$:
\[ j\leq k/2;\quad j_1+j_2+j\leq k \]
where $j_1$ and $j_2$ are two individual charges and the parameter $k$ is the level of the theory.
The Hilbert space of SU(2)$_k$ anyons grows as $\text{dim}(\mathcal{H}_{\text{fusion}})\propto d^N$
where $d=2\cos(\frac{\pi}{k+2})$ is the quantum dimension.  Since the fusion space grows exponentially in the total number of anyons, if one is keeping track of correlations of the walker with the other anyons, the number of degrees of freedom to
keep track of grows very fast with the number of time steps and the calculations become
inefficient. To calculate the probability distribution of the walker for a given time
step, only the calculation of the trace over the fusion degrees of freedom is necessary.
Fortunately, there exists a connection between the trace over the representations
of the braid group $\bra{\Phi_0}B\ket{\Phi_0}$ and link invariants from knot theory
that simplifies calculations of the trace. The trace over the fusion degrees of
freedom can be understood in two equivalent ways: the usual matrix trace over 
the bracket representation of the density matrix of the fusion state,
or the diagrammatic \emph{quantum trace} which is defined by taking the fusion diagram
representation of the state and joining the open ends of the incoming and outgoing charge indices \cite{bondersonsthesis}.
The choice of which indices are joined together determines a tracing scheme for the
braid word, and the tracing corresponds to a closure of a braid in knot theory,
such that the closed braid forms a link. The usual tracing schemes are \emph{plat tracing},
where the neighbouring charges are fused such that their total charge is vacuum,
and \emph{Markov tracing}, where each charge is connected to itself before and after time
evolution.

\begin{figure}[h]
\includegraphics[width=.6\textwidth]{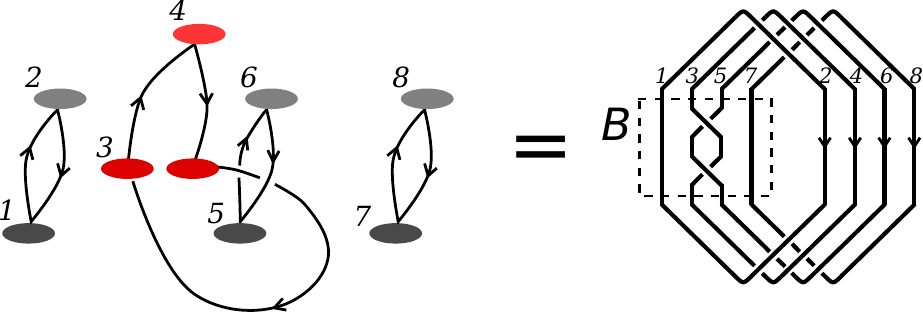}
\caption{A link generated by the anyon worldlines in a quantum walk.
Particle-antiparticle pairs are initially created from the vacuum state and one member of each pair
is dragged out of the system. The walker anyon braids with the background anyons for two steps,
after which each particle is fused back with their antiparticle pairs. Such a protocol implies
that the diagrammatic tracing scheme is Markov.}
\label{fig:aqwlink}
\end{figure}

We choose the initial state of the anyons $\ket{\Phi_0}\bra{\Phi_0}$ such that the tracing corresponds to Markov tracing.
The anyons are initially created from vacuum pairs, after which
one member of each pair is dragged out of the system. In a braid presentation, this corresponds
to having nearest-neighbour pairs with total charge vacuum, followed by a braid word which
moves all left members of the pairs to the left.
The total number of background anyons in an $N$-site walk is $N-1$, but the total Hilbert
space accommodates $2N$ anyons (background anyons, walker anyon, and their antiparticle
pairs). Only half of the anyons participate in the quantum walk and after the two-step evolution
all anyons are fused back together with their antiparticle pairs, see Fig. \ref{fig:aqwlink}.
In this scheme, the expectation value $\bra{\Phi_0}B\ket{\Phi_0}$ corresponds diagrammatically
to the Markov trace over the braid word $B$.

The virtue of using the state $\ket{\Phi_0}$ as the initial fusion state is that
the expectation value $\bra{\Phi_0}B\ket{\Phi_0}$ can now be expressed in terms of the link
invariant called the Kauffman bracket which is a Laurent polynomial in the variable $A$ \cite{kauffman}, denoted $\big\langle L\big\rangle(A)$,
where $L$ is a link, or a collection of tangled loops. This is written formally as
\begin{equation} \label{eqn:expectkauffman}
\bra{\Phi_0}B\ket{\Phi_0}=\frac{\big\langle L\big\rangle(A)}{d^{n-1}}
\end{equation}
where $L=(B^{{\rm Markov}})$ is the link obtained from the braid word $B$ via closure by the initial state $\ket{\Phi_0}$,
$d$ is the quantum dimension of the anyons, $n$ is the number of distinct components (strands) in the link $L$,
and the value of the parameter for SU(2)$_k$ anyons is $A=ie^{-i\pi /2(k+2)}$.
The calculation of the trace over the fusion states thus reduces to evaluation of the Kauffman
bracket of the link that is drawn when the walker anyon moves on a particular path. The evaluation of the
Kauffman bracket is not necessarily more efficient than calculating the matrix trace of the braid
representations, but if one restricts to a model where the walker evolves coherently only for a
finite amount of time steps, it turns out that for arbitrary long walks only a fixed number of links needs to be evaluated
and the quantum walk distribution can be calculated exactly for fairly large number of time steps.

In the usual quantum walk protocol, the quantum speedup occurs because the position of the walker
becomes entangled with the coin. The system evolves coherently and the correlations between the position states
and coin states preserve the memory in the system, hence the system dynamics is highly non-Markovian.
If the walk is subject to decoherence, the correlations between the position and the coin become
degraded and some of the memory in the system is lost. The loss of memory can be modeled using the
so called ``$V^n$ model" \cite{vkmodel}, where the state of the coin is erased and reset on every $n$th
step, for example by doing a measurement on the coin and preparing it over again in the initial state.
After resetting the coin, the system is in a product state and the correlations between the position
and coin are lost.  Remarkably even in the $V^2$ model there is enough coherence left in the system to provide for quadratic speed up over the classical random walk.  Specifically, it was shown in Ref. \cite{vkasymp} that the variance asymptotically scales like $\sigma(t)^2=K_2 t^2+K_3 t$ for some constants $K_2,K_3$.

Here the dynamics of our anyonic walk, Eq. \ref{eqn:aqwevolution1}, is the $V^2$ model where $V=U_SU_F$.  We write the dynamics as a super operator on the density operator $\rho_s(t)$ for the spatial degrees of freedom of the walker:
\[
\begin{array}{rcl}
\rho_{\text{s}}(t+1)&=&\mathcal{E}(\rho_{\text{s}}(t))\\\\
&=&\sum\limits_{f,c}\Big[\big(I\otimes\bra{f}\otimes\bra{c}\big)\big(U_{\text S}U_{\text F}\big)^2\Big]\;
\rho_{\text{s}}(t)\otimes\ket{\Phi_0}\bra{\Phi_0}\otimes\ket{c_0}\bra{c_0}\;
\Big[\big((U_{\text S}U_{\text F})^2\big)^\dag\big(I\otimes\ket{f}\otimes\ket{c}\big)\Big]\\\\
&=&\sum\limits_{f,c}\Big[\big(I\otimes\bra{f}\otimes\bra{c}\big)\big(U_{\text S}U_{\text F}\big)^2
\big(I\otimes\ket{\Psi_0}\otimes\ket{c_0}\big)\Big]\; \rho_{\text{s}}(t)\\\\
&&\Big[\big(I\otimes\bra{\Phi_0}\otimes\bra{c_0}\big)\big((U_{\text S}U_{\text F})^2\big)^\dag
\big(I\otimes\ket{f}\otimes\ket{c}\big)\Big]\\\\
&\equiv&\sum\limits_{f,c}E_{fc}\;\rho_{\text{s}}(t)\;E_{fc}^\dagger
\end{array}
\]
where the Kraus generators have been defined as
\begin{equation} \label{eqn:krausgens}
E_{fc}=\big(I\otimes\bra{f}\otimes\bra{c}\big)\; \big(U_{\text S}U_{\text F}\big)^2\;
\big(I\otimes\ket{\Psi_0}\otimes\ket{c_0}\big).
\end{equation}
Using equations (\ref{eqn:aqwevolution1})--(\ref{eqn:aqwevolution4}) we find for the double step operator
\begin{equation} \label{eqn:doublestep}
\begin{array}{rcl}
(U_{\text{S}}U_{\text{F}})^2&=&\sum\limits_{s=0}^{N-1}\Big(\ket{s-2}\bra{s}\otimes b_{s-2}b_{s-1}\otimes P_0HP_0H
\;+\;\ket{s}\bra{s}\otimes b_{s}^2\otimes P_0HP_1H\\\\
&&+\;\ket{s}\bra{s}\otimes b_{s-1}^2\otimes P_1HP_0H\;+\;\ket{s+2}\bra{s}\otimes b_{s+1}b_{s}\otimes P_1HP_1H\Big).
\end{array}
\end{equation}
It is clear from this equation
that the double step operator shifts the coefficients of the density matrix by two rows or columns, or not at all.
By inserting this expression to the superoperator and using the completeness of the fusion and coin bases,
$\sum\limits_f\bra{f}b_s\ket{\Phi_0}\bra{\Phi_0}b_{s'}^\dagger\ket{f}=\bra{\Phi_0}b_{s'}^\dagger b_s\ket{\Phi_0}$
and $\sum\limits_c\bra{c}P_aHP_bH\ket{c_0}\bra{c_0}H^\dagger P_{a'}H^\dagger P_{b'}\ket{c}
=\delta_{a,b'}\bra{c_0}H^\dagger P_{a'}H^\dagger P_aHP_bH\ket{c_0}$, the action of the superoperator on a general element
$\ket{s}\bra{s'}$ of the spatial density matrix can be written as a sum of 7 terms:
\begin{equation} \label{eqn:superoperator}
\begin{array}{rcl}
\mathcal{E}(\ket{s}\bra{s'}) &=&\frac{1}{4}\Big[\ket{s-2} \bra{s^{\prime }-2}\;
\bra{\Phi_0}b_{s^{\prime }-1}^{\dagger}b_{s^{\prime }-2}^{\dagger }b_{s-2}b_{s-1}\ket{\Phi_0}\;
+\;\ket{s-2}\bra{s^{\prime }}\; \bra{\Phi_0}b_{s^{\prime }}^{\dagger 2}b_{s-2}b_{s-1}\ket{\Phi_0} \\\\
&&+\;\ket{s}\bra{s^{\prime }-2}\; \bra{\Phi_0}b_{s^{\prime }-1}^{\dagger }b_{s^{\prime }-2}^{\dagger}b_{s}^{2}\ket{\Phi_0}
+\;\ket{s}\bra{s^{\prime}}\; \Big(\bra{\Phi_0}b_{s^{\prime }}^{\dagger 2}b_{s}^{2}\ket{\Phi_0}\;
+\; \bra{\Phi_0}b_{s^{\prime }-1}^{\dagger 2}b_{s-1}^{2}\ket{\Phi_0}\Big) \\\\
&&-\;\ket{s}\bra{s^{\prime }+2}\; \bra{\Phi_0}b_{s'}^{\dagger}b_{s^{\prime }+1}^{\dagger}b_{s-1}^{2}\ket{\Phi_0}\;
-\;\ket{s+2}\bra{s^{\prime }}\; \bra{\Phi_0}b_{s^{\prime }-1}^{\dagger2}b_{s+1}b_{s}\ket{\Phi_0} \\\\
&&-\;\ket{s+2}\bra{s^{\prime }+2}\; \bra{\Phi_0}b_{s^{\prime }}^{\dagger }b_{s^{\prime }+1}^{\dagger}b_{s+1}b_{s}\ket{\Phi_0} \Big].
\end{array}
\end{equation}
where we have chosen $\ket{c_0}=\ket{0}$ such that $\bra{c_0}H^\dagger P_{a'}H^\dagger P_aHP_bH\ket{c_0}=\pm\frac{1}{4}$.

Equation (\ref{eqn:superoperator}) shows that for any element of the initial spatial density matrix, there are only
seven nonzero elements of the superoperator $\mathcal{E}$. Writing the density matrix in a vectorized form,
the superoperator can be written as a matrix: $\vec{\rho}\,'=\sum_{f,c}\big(E_{fc}\otimes E_{fc}^\dag\big) \vec{\rho}$,
where the superoperator matrix $\sum_{f,c}\big(E_{fc}\otimes E_{fc}^\dag\big)$ is a band matrix with seven diagonal
bands, and the rest of the elements are zero. Starting from the initial position density matrix
$\rho_{\text{s}}(0)=\ket{s_0}\bra{s_0}$, the position density matrix after $t$ time steps is obtained by applying
the superoperator $t$ times:
\begin{equation} \label{eqn:spatialdm}
\rho_{\text{s}}(t)=\mathcal{E}^t\big(\rho_{\text{s}}(0)\big).
\end{equation}

The elements of the superoperator are given by the product of the coin term
$\bra{c_0}H^\dagger P_{a'}H^\dagger P_aHP_bH\ket{c_0}=\pm\frac{1}{4}$ and the fusion term
$\bra{\Phi_0}B(s,s')\ket{\Phi_0}$. The probability distribution at time step $t$
is obtained by starting from an initially localized state $\ket{s_0}\bra{s_0}$ and applying the superoperator $t$ times.
The site probabilities are then given by the diagonal elements of the spatial density matrix. We have calculated
the variance, $\sigma(t)^2=\langle s^2\rangle-\langle s\rangle^2$, of the $V^2$ walk for various choices of the
level $k$ up to $t=100$ iterations of the superoperator, corresponding to 200 quantum walk steps.

\begin{figure}[h]
a)
\includegraphics[width=.45\textwidth]{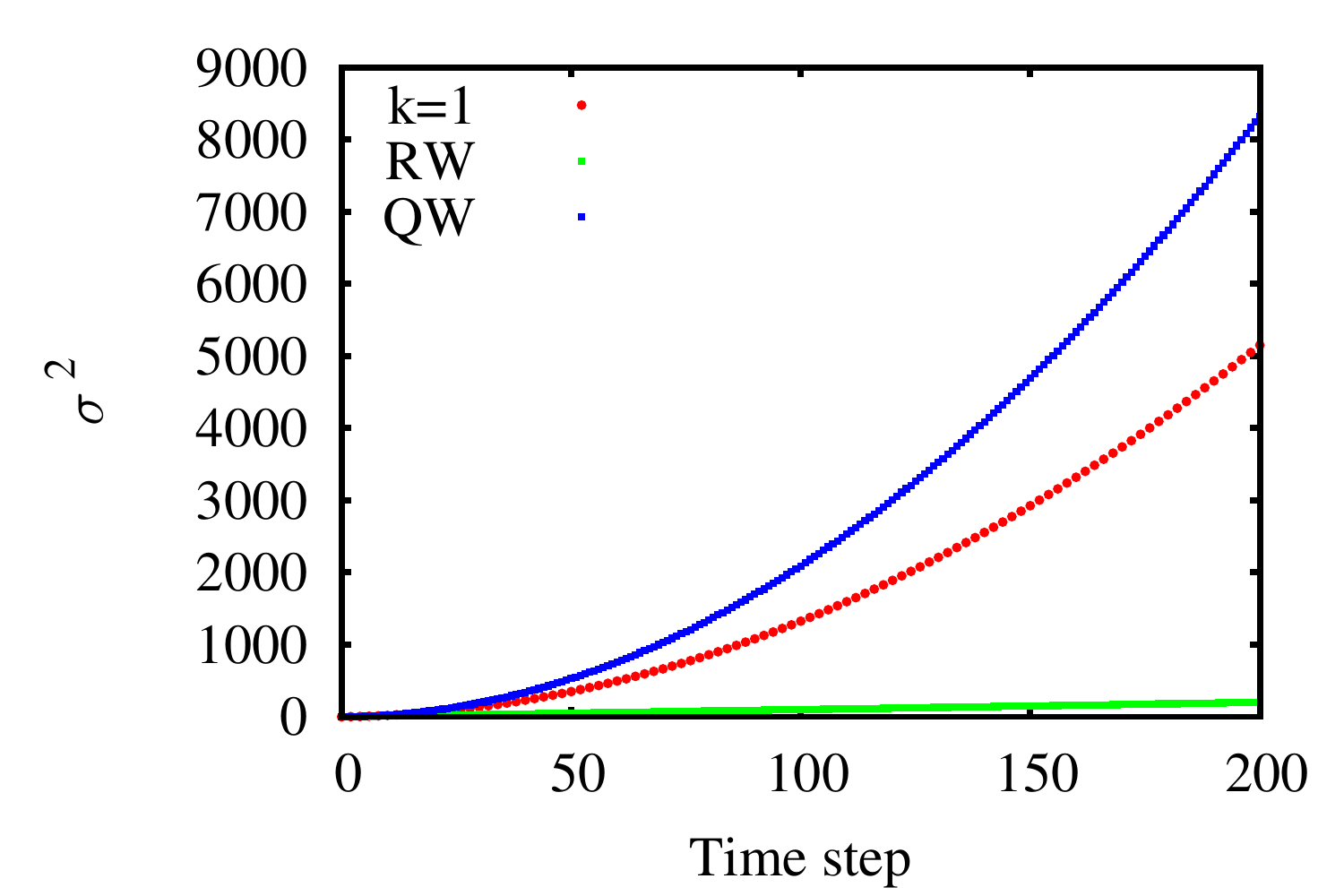}
\quad
b)
\includegraphics[width=.45\textwidth]{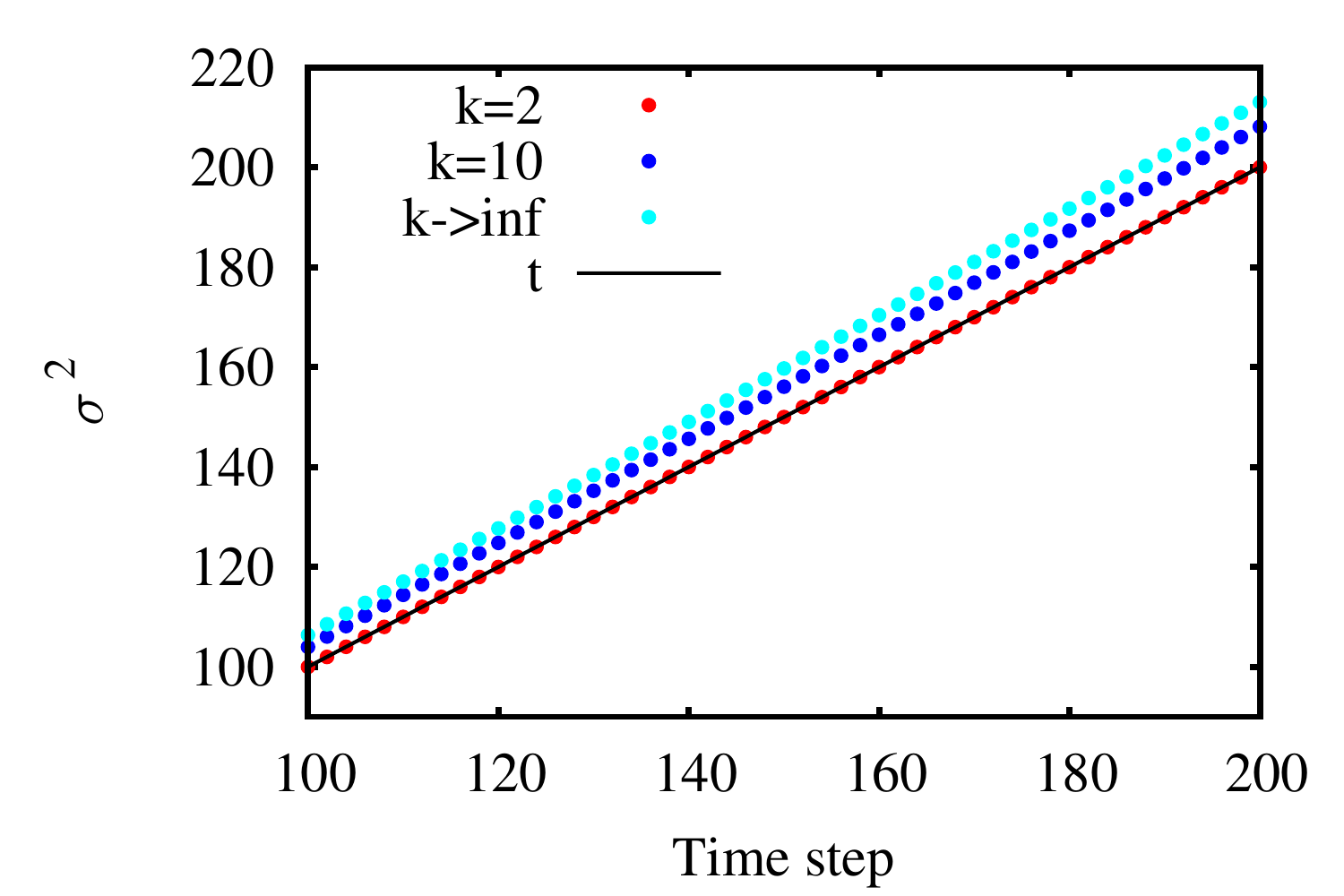}
\caption{The variance $\sigma^2$ of the two-step walk as a function of time $t$ for various anyon models indexed by $k$. The scaling of the sites and time steps is chosen such that one iteration
of the superoperator $\mathcal{E}$ corresponds to two iterations of the quantum walk time evolution operator, and
at each iteration of $\mathcal{E}$ the walker shifts twice. a) Case $k=1$ corresponds to the Abelian walk, which also
coincides with the original $V^2$ quantum walk. Plotted for
comparison are the variance of the classical random walk (RW) and the ordinary quantum walk (QW) with the same initial
coin state and coin flip matrix. b) Non-Abelian SU(2)$_k$ models zoomed to the last 100 time steps. Also plotted is
the linear curve $\sigma^2=t$ which corresponds to the classical random walk.}
\label{fig:variance}
\end{figure}

One interesting feature of the $V^2$ anyonic walk is that the effect of braiding statistics is trivial for
Abelian anyons. If all the stationary anyons are of the same type, the walker always picks the same phase $e^{i\phi/2}$ when
braiding with them, such that
\begin{equation}
\bra{\Phi_0}B\ket{\Phi_0}=e^{-i\phi/2}e^{-i\phi/2}e^{i\phi/2}e^{i\phi/2}=1.
\end{equation}
This is similar to the fully coherent anyonic quantum walk, where the wave function always picks the phase $e^{i\phi t/2}$
during the forward time evolution and phase $e^{-i\phi t/2}$ during backward time evolution, such that the
overall effect is trivial. Another way to think about the Abelian walk is to look at the case $k=1$ in Chern-Simons theory.
There are only two charges $\{0,\frac{1}{2}\}$ in this model, the fusion channels are unambiguous:
$j_1\times j_2=j$, and the quantum dimension is $d=1$, so this model is Abelian. Substituting $A=ie^{-i\pi/6}$ to the values
of the brackets in Table \ref{table:expectvalues} in Appendix \ref{sec:brackets} gives $\bra{\Phi_0}B(s,s')\ket{\Phi_0}=1$ for all
brackets, confirming that the Abelian walk gives the same dynamics as the original $V^2$ quantum walk. The variance
of the $V^2$ walk for Abelian anyons is plotted in Fig. \ref{fig:variance} a). It shows the known fact \cite{vkmodel} that the
original $V^2$ walk propagates slower than the fully coherent quantum walk, but qualitatively the behaviour is still the same:
the variance depends quadratically on time. The best fit for the variance of the $k=1$ walk is given by $\sigma^2=0.125\,t^2+0.75\,t$.

In the non-Abelian case, the construction of the superoperator $\mathcal{E}$ requires evaluation of all eight expectation values
in Eq. (\ref{eqn:superoperator}) for all $s$ and $s'$. The evaluation can be done by using the relation between
the expectation values and the Kauffman bracket in Eq. (\ref{eqn:expectkauffman}). Each braid word consists of
four braid generators, so that each planar representation of the corresponding link has four crossings. A very convenient
feature of the link diagrams is that if $s$ and $s'$ are far apart, the forward and backward braid words act
on disjoint sets of strands, and the Kauffman brackets of all disjoint braids are equal
(see Fig. \ref{fig:links} in Appendix \ref{sec:brackets}). Thus, the calculation
of each expectation value involves only calculating the disjoint case and a small number of cases where $s'$ is
close to $s$. The values of the expectation values are tabulated in Appendix \ref{sec:brackets} for all unique
choices of $s'-s$.

The variances for various choices of $k$ are plotted in Fig. \ref{fig:variance} b). We observe that for $k=2$, the variance follows
exactly the linear line $\sigma^2=t$. It is interesting to note that this behaviour is exactly the same as was
shown for the fully coherent walk with $k=2$ \cite{isingaqw}. For the rest of the values of $k$, there is surprisingly small
difference in the variance, with the slope of the variance slightly increasing as a function of $k$. For $k=3$ and
$k=4$ (not shown in the figure) we find that the slope of the variance is slightly smaller than 1, $k=3$ having the
smallest slope (best fit 0.9877). The highest variance
is obtained when $k\rightarrow\infty$, which corresponds to choice of the parameters $A=i$ and $d=2$, with slope 1.0665.
It should be noted that in the fully coherent anyonic quantum walk model, taking the limit $k\rightarrow\infty$ leads
to the conventional SU(2) algebra, and the braiding generators are a non-Abelian representation of the permutation group.  However when we introduced decoherence in the system, the braid words that contribute to the walk have different structure
than in the fully coherent model. In the fully coherent model, the diagonal elements of the spatial degree of freedom of the walker correspond to a sum over all paths where the walker strand and necessarily all other strands return their initial position
after forward and backward time evolution.  Hence each of these paths is a trivial permutation in the fusion space and has no effect on the diagonal elements of the quantum walk. 
In the $V^2$ model however, the links do not have a dedicated walker strand as can be seen from Fig. \ref{fig:links} in Appendix
\ref{sec:brackets}. The nontrivial character of the $k\rightarrow\infty$ model thus follows from the fact that the links are not identity permutations and hence
expectation values in Table \ref{table:expectvalues} are not equal to 1, as they would be in the fully coherent walk.
The key difference is that only for a highly non-Markovian environment does the $k\rightarrow \infty$ case reduce to the normal
quantum walk for only if one includes memory of all previous steps does the fusion DOF become disentangled with the spatial DOF.

\section{Approximation by circulant matrices}
\label{sec:approx}
In the previous section, we derived the exact superoperator that describes how the spatial density matrix transforms
in a single step of the walk, and obtained numerical results by applying the superoperator to the density matrix repeatedly.
We did not however obtain any expression for the density matrix after evolution by an arbitrary number of steps.
In the following, we approximate the Kraus operators $E_{fc}$ by circulant matrices which have uniform coefficients on
each diagonal, and diagonalize the Kraus operators via Fourier transform to find arbitrary powers of the superoperator
$\mathcal{E}$ in a compact form.

Calling Eqs. (\ref{eqn:krausgens}) and (\ref{eqn:doublestep}) from previous section, the Kraus generators are written explicitly as
\begin{eqnarray}
E_{fc} &=&\sum_{s}\Big[ C_{c}^{00}\;\bra{f}b_{s}b_{s+1}\ket{\Phi_0}\; \ket{s}\bra{s+2}\;
+\;\bra{f}\;(C_{c}^{01}b_{s}^{2}\;+\;C_{c}^{10}b_{s-1}^{2})\;\ket{\Phi_{0}}\; \ket{s}\bra{s} \nonumber \\
&&+\;C_{c}^{11}\;\bra{f}b_{s+1}b_{s}\ket{\Phi_{0}}\; \ket{s+2}\bra{s} \Big] \label{kraus}
\end{eqnarray}
where the coin term has been defined as $C_{c}^{ab}:=\bra{c}P_{a}HP_{b}H\ket{c_0}$.

The matrix expression of the generators is
\begin{equation}
E_{fc}=\left( 
\begin{array}{cccccc}
d_{fc}(0) &  & a_{fc}(0) & \cdots & b_{fc}(N-2) &  \\ 
& d_{fc}(1) &  & a_{fc}(1) &  & b_{fc}(N-1) \\ 
b_{fc}(0) &  & d_{fc}(2) &  & \ddots &  \\ 
& b_{fc}(1) &  & \ddots & \ddots & a_{fc}(N-3) \\ 
a_{fc}(N-2) & \cdots & \ddots &  & \ddots &  \\ 
& a_{fc}(N-1) &  & b_{fc}(N-3) &  & d_{fc}(N-1)
\end{array}
\right) ,  \label{ebm}
\end{equation}
where
\begin{eqnarray}
a_{fc}(s) &=&C_{c}^{00}\bra{f}b_{s}b_{s+1}\ket{\Phi _{0}}, \label{bmatrelem} \\
b_{fc}(s) &=&C_{c}^{11}\bra{f}b_{s+1}b_{s}\ket{\Phi_{0}}, \\
d_{fc}(s) &=&C_{c}^{01}\bra{f}b_{s}^{2}\ket{\Phi_{0}}+C_{c}^{10}\bra{f}b_{s-1}^{2}\ket{\Phi_{0}}.
\end{eqnarray}

Comparing the matrix above with the general circulant matrix given in Eq. (\ref{circularm}) of the
Appendix \ref{sec:circmtx}, we see that although matrix $E_{fc}$
is not a circulant matrix, it can be turned to a circulant matrix 
if the $s$-dependence of parameters $a_{fc}(s)$, $b_{fc}(s)$, $d_{fc}(s)$ is suppressed in some meaningful way, so
that the sequences of these parametrers get approximated by some respective
sequences $\widetilde{a}_{fc}$, $\widetilde{b}_{fc}$ and $\widetilde{d}_{fc}.$ This suppression of the
$s$-dependence also leads to circulant form for $E_{fc}$. The advantage of having the Kraus generators being
circulant matrices stems from the fact that the spectral decomposition problem of such matrices is solved
(see Appendix \ref{sec:circmtx}), and therefore this enables us to express the CP map of our QW and its powers in terms of the
orthogonal eigenprojections of the generators. To this end we introduce for
each $a_{fc}(s),$ $b_{fc}(s)$ and $d_{fc}(s)$ the following
approximation for each $s\in 
\mathbb{Z}
_{N}$
\begin{eqnarray}
a_{fc}(s) &\approx &\widetilde{a}_{fc}(s)=\frac{1}{N}Tr(\widehat{h}^{2}E_{fc}) \label{approx} \\
d_{fc}(s) &\approx &\widetilde{d}_{fc}(s)=\frac{1}{N}Tr(E_{fc}) \\
b_{fc}(s) &\approx &\widetilde{b}_{fc}(s)=\frac{1}{N}Tr(\widehat{h}^{-2}E_{fc})\label{eqn:approx3}
\end{eqnarray}
where the elementary band matrices are defined as $\widehat{h}=\sum_{s\in\mathbb{Z}_{N}}\ket{s}\bra{s+1}$.
This is explicitly written as
\begin{eqnarray}
\widetilde{a}_{fc}&=&\frac{C_{c}^{00}}{N}\sum_{s}\bra{f}b_{s}b_{s+1}\ket{\Phi_{0}}, \label{averaging} \\
\widetilde{d}_{fc}&=&\frac{C_{c}^{01}}{N}\sum_{s}\bra{f}b_{s}^{2}\ket{\Phi_{0}}+\frac{C_{c}^{10}}{N}
\sum_{s}\bra{f}b_{s-1}^{2}\ket{\Phi_{0}}  \\
\widetilde{b}_{fc}&=&\frac{C_{c}^{11}}{N}\sum_{s}\bra{f}b_{s+1}b_{s}\ket{\Phi_{0}},
\end{eqnarray}
i.e. the parameters $a_{fc}(s),$ $b_{fc}(s),$ $d_{fc}(s)$
are approximated by the arithmetic averages of the corresponding $(f,\Phi_{0})$-elements of the respective generators.
Denote $E_{fc}^w$ the matrices obtained from $E_{fc}$ by substituting the $s$-dependent terms with their averages:
\begin{eqnarray}
E_{fc}^w&=&\widetilde{a}_{fc}\sum_{s}\ket{s}\bra{s+2}+\widetilde{d}_{fc}\sum_{s}\ket{s}\bra{s}
+\widetilde{b}_{fc}\sum_{s}\ket{s+2}\bra{s} \\
&=&\widetilde{a}_{fc}\widehat{h}^2+\widetilde{d}_{fc}\widehat{I}+\widetilde{b}_{fc}\widehat{h}^{-2}.
\end{eqnarray}
The map given by the new Kraus generators $E_{fc}^w$ is not trace preserving, so we are seeking for a set of generators
$\widetilde{E}_{fc}$ which satisfy the trace preserving condition
$\sum_{f,c}\widetilde{E}_{fc}^{\dagger}\widetilde{E}_{fc}=\mathbb{I}$.
To this end we need to scale the circulant matrices $E_{fc}^{w}$
by $\sum_{fc}E_{fc}^{w\dagger}E_{fc}^{w}\equiv M$, so the new trace
preserving circulant Kraus generators are defined
$\widetilde{E}_{fc}=(\sum_{f,c}E_{fc}^{w\dagger}E_{fc}^{w})^{-1/2}\:E_{fc}^{w}\equiv\Lambda\:E_{fc}^w$ with $\Lambda=M^{-1/2}$.

A calculation gives
\begin{equation}
M=\kappa_1\:\mathbb{I}+\kappa_2\:\widehat{h}^2+\kappa_2^{\ast}\:\widehat{h}^{-2}
\end{equation}
where
\begin{equation}
\begin{array}{lll}
\kappa_1&=&\sum\limits_{f,c}\big(|\widetilde{a}_{fc}|^2+|\widetilde{d}_{fc}|^2+|\widetilde{b}_{fc}|^2\big)\\
&=& \frac{1}{4}\Big(\frac{1}{N^2}\sum_{j,j'}\bra{\Phi_0}b_{j^{\prime }+1}^{\dagger }b_{j^{\prime}}^{\dagger }b_{j}b_{j+1}\ket{\Phi_0}+\frac{2}{N^2}\sum_{j,j'}\bra{\Phi_0}b_{j^{\prime }}^{\dagger 2}b_{j}^2\ket{\Phi_0} + \frac{1}{N^2}\sum_{j,j'}\bra{\Phi_0}b_{j^{\prime }}^{\dagger }b_{j^{\prime}+1}^{\dagger }b_{j+1}b_{j}\ket{\Phi_0}\Big) \\\\
\kappa_2&=&\frac{1}{4} \Big(\frac{1}{N^2}\sum_{j,j'}\bra{\Phi_0}b_{j^{\prime }}^{\dagger}b_{j^{\prime}+1}^{\dagger}b_{j-1}^2\ket{\Phi_0}+\frac{1}{N^2}\sum_{j,j'}\bra{\Phi_0}b_{j^{\prime }}^{\dagger 2} b_{j}b_{j+1}\ket{\Phi_0}\Big).\\
\end{array}
\label{kappas}
\end{equation}
The matrix $M$ is a circulant band matrix which can be diagonalized via discrete Fourier transform $F$ (see Appendix \ref{sec:circmtx}):
\[
F^\dag MF=\sum_l\big(\kappa_1+\kappa_2\omega^{2l}+\kappa_2^{\ast}\omega^{-2l}\big)\ket{l}\bra{l}.
\]
The matrix $F^\dag MF$ is a diagonal matrix with all diagonal values nonzero, so its inverse exists. The normalization operator $\Lambda$ can now be defined as
$\Lambda=F(F^\dag MF)^{-1/2}F^\dag$, where $(F^\dag MF)^{-1/2}=\sum_l\big(\kappa_1+\kappa_2\omega^{2l}+\kappa_2^{\ast}\omega^{-2l}\big)^{-1/2}\ket{l}\bra{l}$.
A simple check verifies that $\Lambda^2M=I$ and $\Lambda=M^{-1/2}$ as desired.

Unfortunately, the normalized generators $\widetilde{E}_{fc}=\Lambda E_{fc}$ contain all even powers of the matrices $\hat{h}$ for general values of the
Chern-Simons parameter $k$ and do not admit a helpful form. However, for the special values $k=2,4$ there is a significant simplification.  Notice that each of the terms in expressions for $\kappa_1,\kappa_2$ in Eq. (\ref{kappas}) is an averaged Kauffman bracket.  For large $N$ most of these brackets will correspond to disjoint links, i.e. the links formed by two step walks involving the strands located at $j$ and $j^{\prime}$ do not entangle.  As shown in Table \ref{table:expectvalues}, this always occurs if $|j^{\prime}-j|>3$.  If we approximate the average by the value of the bracket for these disjoint links then

\begin{equation}
\begin{array}{lll}
\kappa_1&=&\frac{1}{32}\Big(6\cos(\frac{2\pi}{k+2})+4\cos(\frac{4\pi}{k+2})+2\cos(\frac{6\pi}{k+2})+5\Big)\sec^4(\frac{\pi}{k+2}) +O(1/N)  \\\\
\kappa_2&=&\frac{1}{8}\Big(\cos(\frac{4\pi}{k+2})-\cos(\frac{2\pi}{k+2})+1\Big)\sec^2(\frac{\pi}{k+2})\Big)+O(1/N).
\end{array}
\end{equation}
where the error is of order $1/N$.  Thus at the special values $k=2,4$ we have $\kappa_2\approx 0$ and the normalization operator becomes a
scalar multiple of the identity: $\Lambda=\kappa_1^{-1/2}\:\mathbb{I}$.  Henceforth we focus on the case $k=2$ since the behaviour of $k=4$ is quite similar. The Kraus generators are given by
\begin{equation}
\widetilde{E}_{fc}=\kappa_1^{-1/2}\big(\widetilde{a}_{fc}\widehat{h}^2+\widetilde{d}_{fc}\widehat{I}+\widetilde{b}_{fc}\widehat{h}^{-2}\big).
\end{equation}
The elementary circular matrix $\widehat{h}$ can now be diagonalized as $F^\dag\widehat{h}F=\widehat{g}=\sum_{n\in\mathbb{Z}_N}\omega^n\ket{n}\bra{n}$
where $\omega=e^{2\pi i/N}$. This allows to write the Kraus generators in a diagonal form
\begin{eqnarray}
\widetilde{E}_{fc}&=&\kappa_1^{-1/2}F\big(\widetilde{a}_{fc}\widehat{g}^2+\widetilde{d}_{fc}\mathbb{I}
+\widetilde{b}_{fc}\widehat{g}^{-2}\big)F^\dag\\
&\equiv&\sum_{k\in\mathbb{Z}_N}\lambda_{fc}(k)P_{f_k}
\end{eqnarray}
with $\lambda_{fc}(k)=\kappa_1^{-1/2}\big(\widetilde{a}_{fc}\omega^{2k}+\widetilde{d}_{fc}+\widetilde{b}_{fc}\omega^{-2k}\big)$
and $P_{f_k}=F\ket{k}\bra{k}F^\dag$. The action of the superoperator $\widetilde{\mathcal{E}}$ on a spatial density matrix $\rho_{\text{S}}$ is then given by
\begin{equation}
\widetilde{\mathcal{E}}(\rho_{\text{S}})=\sum_{f,c}\widetilde{E}_{fc}\:\rho_{\text{S}}\:\widetilde{E}_{fc}^\dag
=\sum_{f,c}\sum_{k,l\in\mathbb{Z}_N}\lambda_{fc}(k)\lambda_{fc}^*(l)P_{f_k}\rho_{\text{S}}P_{f_l}^\dag
\end{equation}
and for $t$ time steps
\begin{equation}
\widetilde{\mathcal{E}}^t(\rho_{\text{S}})=\sum_{k,l\in\mathbb{Z}_N}\prod_{i=1}^t\sum_{f_i,c_i}
\big(\lambda_{f_ic_i}(k)\lambda_{f_ic_i}^*(l)\big)P_{f_k}\rho_{\text{S}}P_{f_l}^\dag.
\end{equation}

Let's use this compact form to calculate the diagonal probability distribution $p(s,t)$ at time $t$:
$p(s,t)=\bra{s} \widetilde{\mathcal{E}}^{t}(\rho_{\text{S}}(0))\ket{s}$.  Let's assume that
the walker is initialized in the position eigenstate $\ket{s_0}$, that the size of the periodic lattice is $N$,
and that during the 2-step walk the coin is always reinitialized to state $\ket{c=0}$. We find for the index $k=2$:
\begin{equation} \label{eqn:distk2}
\begin{array}{rcl}
p_{\text{Ising}}(s,t)&=&\langle s | \widetilde{\mathcal{E}}^{t}(|s_0\rangle\langle s_0|)|s\rangle\\
&=&\frac{1}{2^tN^2}\sum_{r,l\in\mathbb{Z}_N}\omega^{(s-s_0)(r-l)}
(\omega^{2(r-l)}+\omega^{-2(r-l)})^t\\
&=&\frac{1}{2^tN^2}\sum_{m=0}^t\binom{t}{m}\sum_{r,l\in\mathbb{Z}_N}\omega^{(s-s_0)(r-l)}
\omega^{2m(r-l)}\omega^{-2(t-m)(r-l)}\\
&=&\frac{1}{2^tN^2}\sum_{m=0}^t\binom{t}{m}\sum_{r\in\mathbb{Z}_N}\omega^{r(s-s_0+4m-2t)}\sum_{l\in\mathbb{Z}_N}\omega^{-l (s-s_0+4m-2t)}\\
&=&\frac{1}{2^tN^2}\sum_{m=0}^t\binom{t}{m}(N\delta_{s-s_0+4m-2t,0})^2\\
&=&\frac{1}{2^t}\binom{t}{\frac{2t-(s-s_0)}{4}}\\
\end{array}
\end{equation}
This is the binomial distribution where the range of the sites is $s\in[-2t,2t]$ and the probabilities are nonzero only for
$s=s_0+4n,\:n\in\mathbb{Z}$, i.e. the $V^2$ model with $k=2$. Ising anyon walkers therefore have the same probability distribution as the classical random walk where every step moves two units to the right or left and the variance, scaled so that each two steps move takes place over two time intervals, is $\sigma_{\rm Ising}^2(t)=t$.

\section{Conclusions}

We have considered a lossy anyonic quantum walk protocol where the entanglement between the spatial modes of
the walker and its environment, the coin and fusion space, is lost on every second step. We calculated the time evolution
of the exact probability distributions for various values of the parameter $k$ in Chern-Simons theory.
The case $k=1$ corresponds to the Abelian anyonic quantum walk and it
was found to be equal to the standard $V^2$ quantum walk with trivial exchange statistics, the variance of which
has a leading term proportional to $t^2$. Cases $k\geq2$ are non-Abelian anyon models, for which the variance
grows linearly as a function of time for all tested values of $k$ in the time scales of up
to 100 iterations of the superoperator.
By approximating the Kraus generators
of the superoperator by circulant matrices, the generators can be diagonalized for $k=2,4$. This allows for a compact
expression for the probability distribution  after arbitrary number
of iterations of the superoperator. The expression for $k=2$ is the binomial distribution which is equal to the
classical random walk distribution. Thus the $V^2$ anyonic quantum walk with Ising anyons has the same behaviour
as the fully coherent walk, which was also shown to have a linear variance with coefficient 1 \cite{isingaqw}.

In the fully coherent walk, the slowdown of the walker propagation can be explained by decoherence: the fusion Hilbert
space of non-Abelian anyons acts like an environment to the walker+coin system, degrading the quantum correlations between
the spatial and coin modes which are the origin of the quantum speedup. It was conjectured in Ref. \cite{isingaqw}
that the slowdown happens when the parameter $k$ is much smaller than the number of time steps $t$, $k\ll t$, ie. for any
finite $k$, the walk behaves classically in the large time limit. In the $V^2$ model presented here the results support that conjecture.  There are two kinds of decoherence
mechanisms at work, one due to the fusion space of the anyons and the other because the entanglement between the spatial
modes and the coin is lost. The loss of quantum correlations  on every second step does not change the qualitative behaviour
when the braiding statistics is Abelian, but when the additional effect of the fusion space comes into play for non-Abelian
anyons, the spreading velocity changes to diffusive for all values of $k$ for long enough times (in fact after fewer than 100 steps).

The $V^2$ model can be easily generalized to $V^n$ models where the walk evolves coherently for $n$ time steps instead of two.
Evaluation of higher number of time steps becomes increasingly hard, because the number of different Kauffman brackets that need to be computed increases as the walker is allowed to do more steps between the tracing operation.
One might expect that for large $n$, the walker propagates diffusively for $n$ steps for small $k$ and ballistically for large $k$,
before the tracing is carried out. However, as we have shown, the variance of the anyonic $V^2$ walk is linear even when $k\gg 2$,
so we expect that in the long time limit the variance is linear for all $V^n$ models, although the walker might spread ballistically
in the initial stage of the walk.





It is known that spatial randomness in the coin operator can lead to localization  of the walker wave packet \cite{localization}, ie. the probability
to stay in the initial position becomes very high and the probability falls off exponentially away from the initial position.
In the anyonic quantum walk setup, spatial randomness can also occur if the occupation numbers of the background anyons
fluctuate randomly. These fluctuations might occur when thermal excitations create particle-antiparticle pairs from vacuum, for example.
It is interesting to note that if the occupation numbers of the islands are not uniform, the Abelian anyons can pick up different
phases during forward and backward time evolution, and the effect of the braiding statistics becomes nontrivial. We have calculated
evolution of the $V^2$ anyonic walk with random fillings of Abelian anyons on the islands, but observed diffusive spreading (no localization).  Localization is a quantum phenomenon that requires interference of the probability amplitudes to occur.  Interestingly our results show that while the memory loss in the $V^2$ model preserves enough coherence in the Abelian walk to provide for quadratic speed up without disorder, it does not provide enough to give localization with disorder.  Rather the competition of the two effects yields classical behaviour.

\section*{Acknowledgements}

One of us (D.E.) is grateful to the Department of Physics and Astronomy, Macquarie University for hospitality during a sabbatical stay during which this work was initiated.

\appendix

\section{Braid group matrix elements} \label{sec:brackets}

\begin{figure}[h]
a)
\includegraphics[height=.3\textwidth]{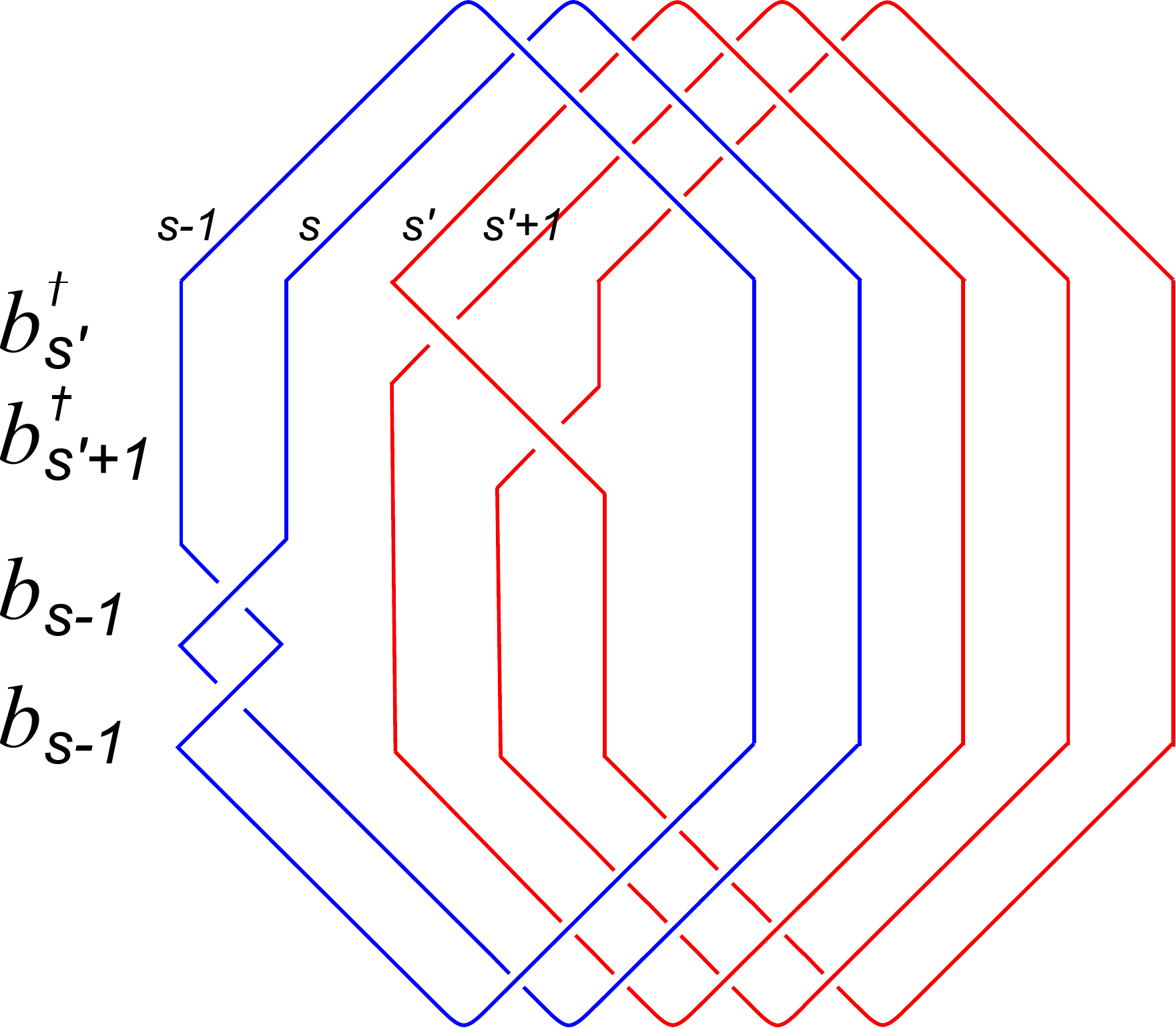}
\hspace{.05\textwidth}
b)
\includegraphics[height=.3\textwidth]{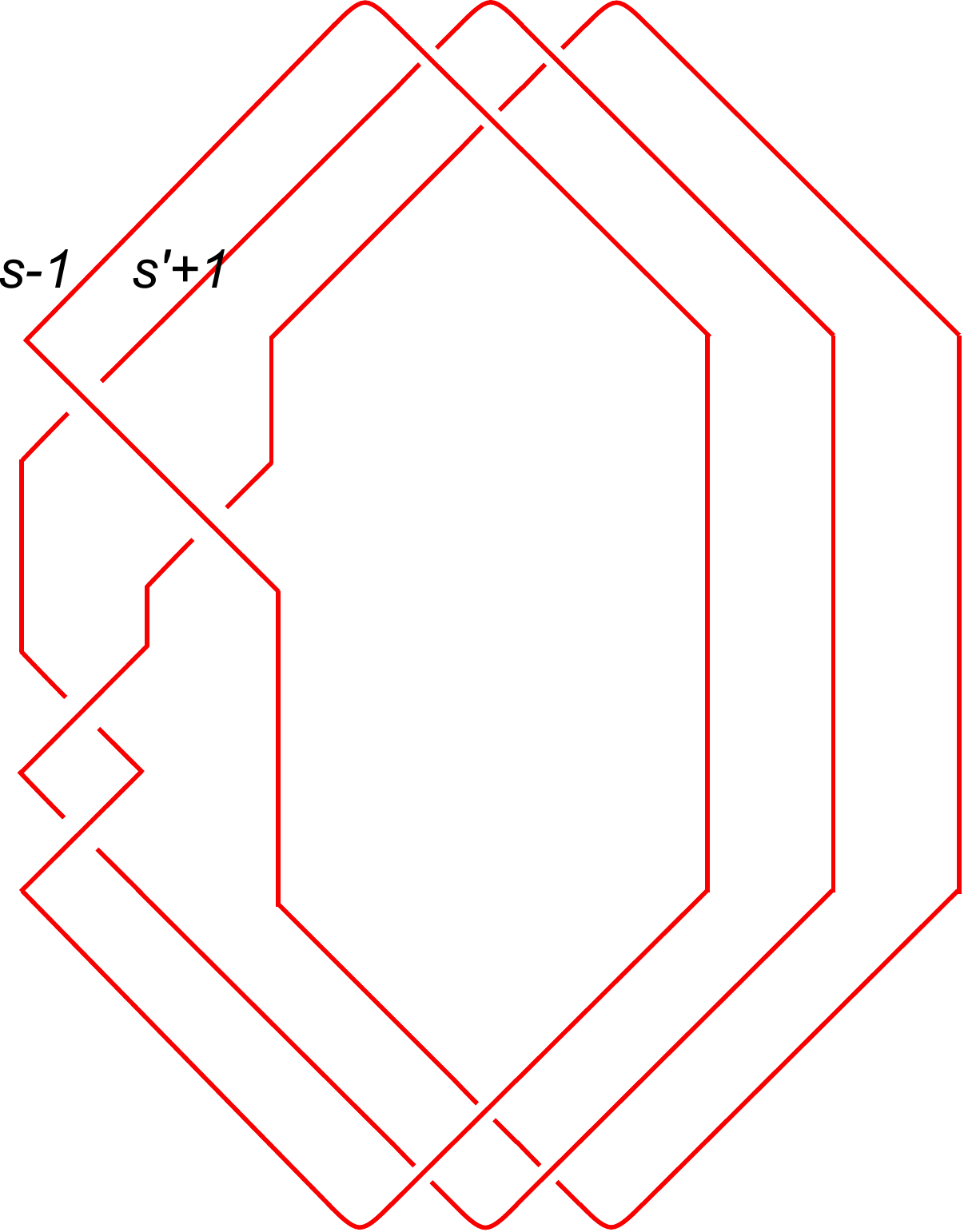}
\caption{Two links corresponding to the expectation value
$\bra{\Phi_0} b_{s^{\prime }}^{\dagger}b_{s^{\prime }+1}^{\dagger }b_{s-1}^{2}\ket{\Phi_0}$.
a) Case $s'-s=1$. The forward braids $b_{s-1}^{2}$ and the backward braids
$b_{s^{\prime }}^{\dagger}b_{s^{\prime }+1}^{\dagger }$ act on separate sets of strands, therefore the links
corresponding to the forward and backward braids are disjoint. The links corresponding to $s'-s\leq-4,\,s'-s\geq1$
are all disjoint and the values of the Kauffman brackets are the same. b) Case $s'-s=-1$. The forward and
backward braids now form a joint link.
}
\label{fig:links}
\end{figure}

The construction of the superoperator $\mathcal{E}$ requires calculating the eight expectation
values in Eq. (\ref{eqn:superoperator}) for all values of $s$ and $s'$. The expectation values are related
to the Kauffman bracket via Eq. (\ref{eqn:expectkauffman}). Kauffman bracket polynomial
$\big\langle L\big\rangle(A)$ in the variable $A$ \cite{kauffman} is a link invariant for framed,
unoriented links. The defining properties of the Kauffman bracket are given by the uncrossing relation
and two relations for removing loops from the bracket:
\begin{equation} \label{eqn:kbracket1}
\Big\langle\raisebox{-.1em}{\includegraphics[width=1em]{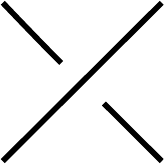}}\Big\rangle
=A\Big\langle\raisebox{-.2em}{\includegraphics[height=1.2em]{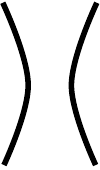}}\Big\rangle
+A^{-1}\Big\langle\includegraphics[angle=90,width=1.2em]{identitycrossing}\Big\rangle
\end{equation}
\begin{equation} \label{eqn:kbracket2}
\big\langle L\cup\bigcirc\big\rangle=-(A^2+A^{-2})\big\langle L\big\rangle
\end{equation}
\begin{equation} \label{eqn:kbracket3}
\big\langle\bigcirc\big\rangle=1.
\end{equation}
The value of the Kauffman bracket for a specific braid word can be calculated by forming the link diagram that
corresponds to the braid presentation and Markov closure of the word. The unique link diagrams for
one braid word are drawn for illustration in Fig. \ref{fig:links}. The value of the bracket polynomial
is then obtained by applying the relations (\ref{eqn:kbracket1}) and (\ref{eqn:kbracket2}) to the link
diagram, removing all crossings and loops in the diagram except one. The bracket of a single loop is trivial,
so the remaining coefficient is the value of the bracket polynomial.

For SU(2)$_k$ anyons, the value of the parameter is $A=ie^{-i\pi /2(k+2)}$, and the quantum dimension
satisfies $d=-(A^{2}+A^{-2})=2\cos(\frac{\pi}{k+2})$. For $s'-s$ large, the forward and backward braid words never touch the same strands,
the links corresponding to forward and back evolution are disjoint, and the value of the Kauffman bracket polynomial
is equal for all $s$ and $s'$. Thus, the calculation of a general element $\bra{\Phi_0}B(s,s')\ket{\Phi_0}$
involves only the calculation of the disjoint element and a few cases where $s'-s$ is small. The unique values
of five of the expectation values are given in Table \ref{table:expectvalues}.
For the remaining elements, we have
\[
\bra{\Phi_0} b_{s^{\prime }-1}^{\dagger 2}b_{s-1}^{2}\ket{\Phi_0}
=\bra{\Phi_0} b_{s^{\prime }}^{\dagger 2}b_{s}^{2}\ket{\Phi_0}
\]
and the conjugate transpose elements
\[
\bra{\Phi_0}b_{s^{\prime }-1}^{\dagger }b_{s^{\prime}-2}^{\dagger }b_{s}^{2}\ket{\Phi_0}
=\big(\bra{\Phi_0} b_{s}^{\dagger 2}b_{s^{\prime }-2}b_{s^{\prime }-1}\ket{\Phi_0}\big)^*
\]
and
\[
\bra{\Phi_0}b_{s^{\prime}-1}^{\dagger 2}b_{s+1}b_{s}\ket{\Phi_0}
=\big(\bra{\Phi_0} b_{s}^{\dagger}b_{s+1}^{\dagger }b_{s^{\prime }-1}^{2}\ket{\Phi_0}\big)^*.
\]

\begin{table}
\begin{tabular}{|l||l|l|}
\hline
$s\prime -s$ & $\bra{\Phi_0} b_{s^{\prime }-1}^{\dagger }b_{s^{\prime}-2}^{\dagger }b_{s-2}b_{s-1}\ket{\Phi_0} $
& $\bra{\Phi_0} b_{s^{\prime}}^{\dagger }b_{s^{\prime }+1}^{\dagger }b_{s+1}b_{s}\ket{\Phi_0} $ \\ \hhline{|=::=|=|}
$-2$ &
$d^{-4}$
& $d^{-4}$ \\ \hline
$-1$ &
$d^{-2}$
& $d^{-2}$ \\ \hline
$0$ & 1 & 1 \\ \hline
$1$ &
$d^{-2}$
& $d^{-2}$ \\ \hline
$2$ &
$d^{-4}$ & $d^{-4}$ \\ \hline
$\leq-3$,\; $\geq 3$ &
$d^{-4}$
& $d^{-4}$ \\ \hline
\end{tabular}
\vspace{3ex}\\
\begin{tabular}{|l||l|l|}
\hline
$s\prime -s$ & $\bra{\Phi_0}b_{s^{\prime }}^{\dagger 2}b_{s-2}b_{s-1}\ket{\Phi_0}$
& $\bra{\Phi_0} b_{s^{\prime }}^{\dagger}b_{s^{\prime }+1}^{\dagger }b_{s-1}^{2}\ket{\Phi_0}$ \\ \hhline{|=::=|=|}
$-3$ &
$-A^6(A^4+A^{-4})/d^3$
& $-A^{-6}(A^4+A^{-4})/d^3$ \\ \hline
$-2$ &
$d^{-2}$
& $d^{-2}$ \\ \hline
$-1$ &
$d^{-2}$
& $d^{-2}$ \\ \hline
$0$ &
$-A^6(A^4+A^{-4})/d^3$
& $-A^{-6}(A^4+A^{-4})/d^3$ \\ \hline
$\leq-4$,\; $\geq1$ &
$-A^6(A^4+A^{-4})/d^3$
& $-A^{-6}(A^4+A^{-4})/d^3$ \\ \hline
\end{tabular}
\vspace{3ex}\\
\begin{tabular}{|l||l|}
\hline
$s\prime -s$ & $\bra{\Phi_0}b_{s^{\prime }}^{\dagger 2}b_{s}^{2}\ket{\Phi_0} $ \\ \hhline{|=::=|}
$-1$ &
$(A^4+A^{-4})^2/d^2$
\\ \hline
$0$ & 1 \\ \hline
$1$ &
$(A^4+A^{-4})^2/d^2$
\\ \hline
$\leq-2$,\; $\geq2$ &
$(A^4+A^{-4})^2/d^2$ \\ \hline
\end{tabular}
\caption{The values of the braid group elements involved in the evolution CP map.}
\label{table:expectvalues}
\end{table}


\section{Circulant matrices and CP map generators} \label{sec:circmtx}

A matrix $C=(c_{ij})$ of order $n$ is called a circulant matrix \cite{circular} if $%
c_{ij}=a_{i-j(\mod n)}$. The entries of the first column $a\equiv
(a_{0},a_{1},...,a_{n-1})$ determine the entire circulant matrix $\bigskip $%
\ which we denote $C_{n}=circ(a)=circ(a_{0},a_{1},...,\alpha _{n-1}),$ and
in matrix form it reads 
\begin{equation}
\bigskip C_{n}=\left( 
\begin{array}{ccccc}
a_{0} & a_{n-1} & a_{n-2} & \cdots & a_{1} \\ 
a_{1} & a_{0} & a_{n-1} & \cdots & a_{2} \\ 
a_{2} &  & a_{0} & \cdots & a_{3} \\ 
\vdots & \vdots & \vdots & \ddots & \vdots \\ 
a_{n-1} & a_{n-2} & a_{n-3} & \cdots & a_{0}%
\end{array}%
\right) .  \label{circularm}
\end{equation}
\ 

Alternatively circulant matrices are written in terms of elementary circulant matrix
$\widehat{h}=\sum_{m\in \{0,1,...,n-1\}}\ket{m}\bra{m+1}=circ(0,1,...,0)$, as
$C_{n}=circ(a_{0},a_{1},...,\alpha _{n-1})=p(\widehat{h})$ where $p$ is the polynomial
$p(z):=a_{0}+a_{1}z+\cdots a_{n-1}z^{n-1}$. Matrix $\widehat{h}$ is diagonalized by the finite Fourier transform
unitary matrix $F$, with elements $F_{ab}=\frac{1}{\sqrt{n}}\omega ^{ab}$, $\omega =e^{i2\pi /n}$, as
$F^{\dagger }\widehat{h}F=\widehat{g}$, where $\widehat{g}=diag(1,\omega,...,\omega ^{n-1})$. Any circulant
matrix is then canonically decomposed as
$C_{n}=circ(a_{0},a_{1},...,\alpha _{n-1})=F^{\dagger }diag(p(1),p(\omega),...,p(\omega ^{n-1}))F.$ A banded
circulant matrix is a circulant matrix for which only a connected subset in the sequence $a=(a_{j})_{j=0}^{n-1}$ is non
zero. After introducing the approximations issued in Eqs. (\ref{approx})--(\ref{eqn:approx3}) the generators
$E_{fc}$ of QW's CP map become banded circulant matrices i.e.
$\widetilde{E}_{fc}=circ(a_{0}=\widetilde{d}_{fc},a_{1}=0,a_{2}=\widetilde{b}_{fc},...,
a_{n-2}=\widetilde{a}_{fc},a_{n-1}=0).$

Circulant matrix $\ \widehat{h}=\sum_{s\in 
\mathbb{Z}
_{N}}\left\vert s\right\rangle \left\langle s+1\right\vert ,$ and its
inverse \ $\widehat{h}^{-1}$ 
\begin{eqnarray}
\widehat{h} &=&\sum_{n}\left\vert n+1\right\rangle \left\langle n\right\vert
=\left( 
\begin{array}{ccccc}
0 &  &  &  & 1 \\ 
1 & 0 &  &  &  \\ 
& 1 & 0 &  &  \\ 
&  & 1 & \ddots &  \\ 
&  &  & 1 & 0%
\end{array}%
\right) , \\
\widehat{h}^{-1} &=&\sum_{n}\left\vert n-1\right\rangle \left\langle
n\right\vert =\left( 
\begin{array}{ccccc}
0 & 1 &  &  &  \\ 
& 0 & 1 &  &  \\ 
&  & 0 & 1 &  \\ 
&  &  & \ddots & 1 \\ 
1 &  &  &  & 0%
\end{array}%
\right) ,
\end{eqnarray}
generate the abelian group $\{\widehat{h}^{a}\}_{a\in 
\mathbb{Z}
_{N}}\simeq
\mathbb{Z}
_{N},\ $share the property\ $\ \widehat{h}^{\dagger }=\widehat{h}%
^{-1}=\sum_{s\in 
\mathbb{Z}
_{N}}\left\vert s+1\right\rangle \left\langle s\right\vert ,$ where $%
\widehat{h}^{2}=\mathbf{1}.$

An important tool used in Sec. \ref{sec:approx} to approximate the $V^2$ quantum walk model is the optimal circulant of a matrix.  For
an arbitrary square matrix $D=(d_{jk}),i,k=0,1,...,N-1,$ we choose a
suitable circulant matrix $C.$ For the construction of $C$ it has been
recommended \cite{berg}, (after preliminary transformations such as changing
the order of the rows and multiplying them with suitable constants), to
apply the following: by imposing periodicity i.e. $d_{j+N,k}=d_{j,k},$ form
the arithmetic averages of the elements along the diagonals i.e. $a_{j}=%
\frac{1}{N}\sum_{k=0}^{N-1}d_{j+k,k},$ and construct in this way the
circulant $C=circ(a_{0},a_{1},...ma_{N-1}).$

On the other hand an optimal circulant approximation of some Toeplitz matrix 
$T$ (see e.g. Strang's suggestion in \cite{strang}), have been constructed
by determing the nearest circulant with respect to Frobenius matrix norm
for matrix $T$, i.e. by solving the optimization problem min$%
_{C:circulant}\left\Vert T-C\right\Vert _{F}$ \cite{chan}. The resulting
circulant coincides with the one obtained by the method of arithmetic
averages \cite{berg}, applied to $T$. In the quantum walk context the method of
arithmetic averages is been applied to the Kraus generators $E_{fc}$ that
are approximated by the optimal circulant $\widetilde{E}_{fc}.$

\end{document}